\documentclass[aps,prl,10pt,showpacs,notitlepage,nofootinbib,superscriptaddress,floatfix,showkeys,twocolumn,preprintnumbers]{revtex4-1}
\usepackage{amsmath,amssymb}
\usepackage{microtype}
\usepackage{graphicx}
\usepackage{bm}
\usepackage{epsfig}
\usepackage{psfrag}
\usepackage{ulem}
\usepackage{color}
\usepackage[dvipsnames]{xcolor}
\usepackage[section]{placeins}
\usepackage{comment}
\usepackage{physics}
\usepackage{siunitx}
\usepackage{booktabs}
\usepackage{xcolor}
\usepackage[colorlinks=true,
            linkcolor=black, 
            citecolor=black, 
            urlcolor=black, 
            ]{hyperref}
\usepackage{orcidlink}
            
\AtBeginDocument{\RenewCommandCopy\qty\SI}

\allowdisplaybreaks[1]

\def\Eein{E_{e}}
\def\Epin{E_{\gamma}}
\def\Epout{E}
\def\EpinERF{E^{\prime}_{\gamma}}
\def\EpoutERF{E^{\prime}}

\begin{document}

\preprint{KCL-PH-TH/2026-11}

\title{A Solar Probe of Dark Matter Decay in the Galaxy}

\author{Maximillian~Detering\,\orcidlink{0009-0001-1408-8192}}\email{maximillian.detering@kcl.ac.uk}
\affiliation{Physics Department, King’s College London, Strand, London, WC2R 2LS, United Kingdom}

\author{Shyam~Balaji\,\orcidlink{0000-0002-5364-2109}}
\email{shyam.balaji@kcl.ac.uk}
\affiliation{Physics Department, King’s College London, Strand, London, WC2R 2LS, United Kingdom}

\smallskip
\begin{abstract}
Dark matter (DM) particles decaying in the Galactic halo can inject energetic $e^\pm$ that inverse-Compton scatter (ICS) solar photons into $\gamma$-rays, producing a diffuse and extended halo of emission around the Sun. We present the first quantitative study of this signal as an indirect probe of decaying DM. The intense solar photon field in the inner heliosphere amplifies the inverse-Compton emissivity by many orders of magnitude relative to the interstellar radiation field, making the Sun an unusually sensitive local converter of sunlight into $\gamma$-rays via scattering with injected $e^\pm$. Using 15 years of {\it Fermi}-LAT solar-halo data, we derive stringent limits on the DM lifetime for 10~GeV--10~TeV masses at the level of $\tau_\chi \sim 10^{27}\,\mathrm{s}$ in leptonic decay channels. The predicted surface brightness rises steeply toward the Sun, and the $\gamma$-ray flux falls off at high energy due to Klein--Nishina suppression. Solar ICS $\gamma$-rays measured with degree scale angular resolution therefore provide a novel and complementary probe of DM decays, adding a local $\gamma$-ray search channel that is systematically distinct from both Galactic diffuse analyses and direct charged-particle measurements.

\end{abstract}
\maketitle


\section{Introduction}
Indirect searches for particle dark matter (DM) typically exploit Standard Model messengers produced by annihilation or decay, with $\gamma$-rays providing a particularly powerful handle due to their directional information and comparatively clean transport.  Most existing $\gamma$-ray searches for decaying DM target large DM columns in the Galactic halo, dwarf galaxies, or the extragalactic sky, and are limited by diffuse backgrounds and modeling systematics that grow as one pushes toward fainter signals.  In this work we develop and apply a qualitatively different strategy: we use the Sun as a \textit{local} amplifier that converts DM-decay electrons and positrons into a distinctive, spatially extended inverse-Compton (IC) $\gamma$-ray halo. The key point is simple. The IC $\gamma$-ray yield along any line-of-sight (LOS) scales with the photon column density rather than the DM column. Galactic probes make use of the interstellar radiation field and cosmic microwave background, typically in the off-disk direction to minimize other backgrounds, resulting in a large photon column density. However, the sunlight column density $N$ observed within a few degrees of the Sun is remarkably comparable due to the intense solar photon field.
As a consequence, any relativistic $e^\pm$ population injected by DM decay in the nearby halo is far more efficient at producing observable $\gamma$-rays through inverse-Compton scattering (ICS) in the solar environment than in the general Galactic medium, turning the Sun into a uniquely sensitive and geometrically well-defined target for DM decay.

The physical ingredients that enable this search are well understood. Cosmic-ray electrons and positrons traversing the heliosphere up-scatter optical photons emitted by the solar surface, generating a diffuse $\gamma$-ray halo around the Sun~\cite{Moskalenko:2006ta,Orlando:2006zs,Orlando:2008uk,Zhou:2016ljf,Orlando:2020ezh,Yang:2023res}.  In the high-energy limit where the local $e^\pm$ distribution is approximately homogeneous and isotropic within the heliosphere, the solar photon field falls as $r^{-2}$ with heliocentric distance, yielding a surface brightness that scales roughly as $\theta^{-1}$, where $\theta$ is the elongation angle, after LOS integration. Two effects shape the near-Sun and low-energy behavior: (i) the anisotropic Klein--Nishina kernel, and (ii) solar modulation by the heliospheric magnetic field, which depletes the $e^\pm$ population at low energies and modifies the IC $\gamma$-ray emissivity in the inner heliosphere~\cite{Gleeson:1968zza,Moskalenko:2006ta,Orlando:2008uk,Orlando:2020ezh}. Crucially for DM searches, these same geometrical and kinematic features imprint a characteristic \textit{joint} energy--angle dependence on any DM-induced ICS signal, providing discrimination power against the broader diffuse $\gamma$-ray backgrounds.

A decisive recent development is that the solar ICS halo is now measured with sufficient precision to be repurposed as an indirect probe of DM.  An early \textit{Fermi}-LAT analysis detected the halo using 1.5 years of data, but required severe background-rejection cuts that limited its utility for new-physics searches~\cite{Fermi-LAT:2011nwz}.  More recently, Ref.~\cite{Linden:2025xom} performed the first high-significance measurement over a full solar cycle by developing a dedicated moving-source analysis and a data-driven background model tailored to the Sun. Using 15 years of \textit{Fermi}-LAT data, they robustly detect the ICS halo from 31.6~MeV to 100~GeV and out to $45^\circ$ from the Sun, while mapping its spectrum, morphology, and time dependence with unprecedented statistical power.  These results establish both the existence and the detailed characteristics of the solar halo at precisely the energies and angular scales most relevant for GeV--TeV DM-induced $e^\pm$ populations.

In this work we compute the solar ICS $\gamma$-ray signal from decaying DM in the full heliocentric geometry and confront it with the measured {\it Fermi}-LAT solar-halo fluxes~\cite{Linden:2025xom}. 
We build energy- and annulus-binned signal templates for benchmark decay channels and derive conservative limits while profiling over solar-modulation nuisance parameters. We emphasize leptonic channel such as $\chi\to e^+e^-$ as a benchmark since it yields the hardest injection spectrum and typically the strongest constraints, and we also present results for several other relevant decay channels. We remark that a similar kind of analysis can also be performed for annihilating DM, which we do not conduct here, since the expected constraints are not competitive with the most sensitive existing indirect searches. This is simply due to the relative impact of a local search is mitigated by the LOS integral over quadratic DM density dependence in the source term of injected particles.

Our results demonstrate that solar IC emission provides a new and complementary probe of decaying DM: it is local, exhibits a unique signal morphology, and is constrained by a dataset with empirically measured backgrounds and a validated solar-halo detection.  In particular, the combination of the steep solar photon gradient and the anisotropic scattering kernel yields a distinctive morphology that peaks at degree scales from the solar disk, allowing competitive sensitivity to DM lifetimes in the GeV--TeV mass range.  More broadly, this approach opens a new avenue for indirect detection that leverages heliospheric photon fields as converters of otherwise diffuse leptonic injection.

The remainder of the paper is organized as follows. In Sec.~\ref{sec:Formalism} we present the heliocentric ICS calculation, including the solar photon field, the modulated $e^\pm$ spectra (astrophysical and DM-induced), and the anisotropic scattering kernel. In Sec.~\ref{sec:Analysis} we describe the \textit{Fermi}-LAT solar-halo dataset and our likelihood framework which leverages our signal templates, and derive conservative profile-likelihood limits.  Our main constraints and comparisons to existing $\gamma$-ray bounds are presented in Sec.~\ref{sec:Results}. We conclude in Sec.~\ref{sec:Conclusion} with a discussion of dominant uncertainties and prospects for strengthening solar ICS DM searches with future high-exposure solar observations and further studies of solar modulation.

\section{Solar \texorpdfstring{$\boldsymbol{\gamma}$}{γ}-rays from inverse Compton scattering}
\label{sec:Formalism}
In this section we describe in detail the calculation of the $\gamma$-ray flux from ICS of solar photons with cosmic rays. We work in natural units, i.e. $\hbar = c = 1$.

Generally, all local charged cosmic ray particle fluxes of sufficiently high energy may contribute to the $\gamma$-ray halo around the Sun. However, since the scattering cross section scales with the charged particle mass as $m^{-2}$, the contributions from the lightest stable charged cosmic ray species, electrons and positrons, completely dominate the production of solar halo $\gamma$-rays. 

\subsection{Emissivity}\label{sec:emissivity}
The emissivity $j_\omega$ describes the particle emission rate per unit energy, volume and solid angle. Consider the scattering of two particle fluxes with phase space distributions $\rho_i = n_i(\mathbf{r}) f_i(\mathbf{p_i})$, where $n$ refers to particle number density, and where the momentum-space distribution function is normalized as
\begin{equation}
    \int \frac{\dd^3 p_i}{(2\pi)^3} f_i(\mathbf{p}_i) = 1.
\end{equation}
The differential emissivity takes the general form \cite{weaver_reaction_1976}
\begin{equation}
    j_\omega = n_1 n_2 \int \frac{\dd^3 p_1 f_1(\mathbf{p}_1)}{(2\pi)^3 2E_1} \frac{\dd^3 p_2 f_2(\mathbf{p}_2)}{(2\pi)^3 2E_2} 4 F \frac{\dd \sigma}{\dd E \dd \Omega},
\end{equation}
where $F=\sqrt{(p_1 \cdot p_2)^2 - m_1^2 m_2^2}$ is the M\o ller flux factor and $\dd \sigma / (\dd E \dd \Omega)$ is the differential cross section. We consider the Compton scattering of ultra-relativistic electrons and positrons, where the differential emissivity simplifies to
\begin{equation}
    j_\omega = n_e n_\gamma \int \frac{\dd E_\gamma \dd \Omega_\gamma }{(2\pi)^3} E_\gamma f_\gamma \frac{\dd E_e \dd \Omega_e}{(2\pi)^3} E_e f_e F \frac{\dd \sigma}{\dd E \dd \Omega}.
\end{equation}
Furthermore, with our normalization of $f_i$, we can identify
\begin{equation}
    \frac{\dd \Phi_e}{\dd \Eein} \equiv \Eein^2 \frac{n_e f_e}{(2\pi)^3},
\end{equation}
where $\dd \Phi_e/\dd \Eein$ is the differential $e^{\pm}$ flux per unit energy, area and solid angle. Similarly, we can write the solar photon field density as
\begin{equation}
    \frac{\dd N_\gamma}{\dd \Epin \dd \Omega_\gamma} \equiv \frac{\Epin^2}{(2\pi)^3} n_\gamma f_\gamma  \; .
    \label{eq:differential-solar-photon-density}
\end{equation}
Furthermore, note that the M\o ller flux factor takes the trivial form in the electron rest frame (ERF)
\begin{equation}
    F = m_e \EpinERF,
\end{equation}
where $\EpinERF = \Epin (\Eein / m_e) (1 - \cos\eta)$ and $\eta$ is the photon scattering angle to the incident electron in the solar rest frame (SRF).
With these, the differential emissivity takes the form
\begin{multline}
    j_\omega(\Epout,\mathbf{r},\Omega)
    = \int \dd \Eein \dd \Epin \dd\Omega_e \dd\Omega_\gamma \; \frac{m_e \EpinERF}{\Eein \Epin}\\
    \times \frac{\dd \Phi_e}{\dd \Eein}(\Eein, \Omega_{e})  \frac{\dd N_\gamma}{\dd \Epin \dd \Omega_\gamma}(\Epin, \Omega_{\gamma,\rm in}) \frac{\dd \sigma}{\dd \Epout \dd \Omega} (\Eein, \Epin, \Epout, \eta) \; .
    \label{eq:emissivity-simplified}
\end{multline}

\subsection{Solar photon field}\label{sec:photon-field}
The solar photon field density from Eq.~\eqref{eq:differential-solar-photon-density} is given in terms of the solar surface intensity $I_{\Epin}$ as
\begin{equation}
    \left.\frac{\dd N_\gamma}{\dd \Epin \dd \Omega_{\gamma}}\right\rvert_{\Omega_{\odot}} = \frac{I_{\Epin}}{\Epin} = \frac{\epsilon_\odot B_{\Epin}(T)}{\Epin} = \frac{\Epin^2}{4\pi^3}\frac{1}{\mathrm{e}^{\Epin/T} - 1},
    \label{eq:solar-photon-density}
\end{equation}
where $\epsilon_{\odot} \approx 1$ is the solar blackbody factor and $B_{\Epin}(T)$ is the Planck energy flux density for a blackbody temperature $T \simeq \SI{5778}{K}$, and $\Omega_{\odot}$ denotes the solid angle of the Sun from point $\mathbf{r}$ from its center. The differential photon number density is constant over the opening angle of the Sun, and we can integrate \eqref{eq:solar-photon-density} over the solid angle
\begin{equation}
    \frac{\dd N_\gamma}{\dd \Epin} = \frac{B_{\Epin} (T)}{\Epin} 2 \pi \left( 1 - \sqrt{1 - \frac{R_\odot^2}{r^2}} \right),
    \label{eq:solar-photon-density-integrated}
\end{equation}
where $R_\odot$ is the solar radius and $r$ is the radial distance to the Sun. For large distances to the Sun, \eqref{eq:solar-photon-density-integrated} reduces to the simple inverse square law; the precise form of Eq.~\eqref{eq:solar-photon-density-integrated} is only relevant for ICS $\gamma$-ray emission for angular distances to the solar center $\theta \lesssim 2.5^\circ$.

\subsection{Cosmic-ray electron and positron fluxes}\label{sec:cre-flux}
We model the cosmic-ray electron and positron fluxes through two independent components: a background flux from astrophysical sources, and a component due to the decay of DM particles within the Galaxy.

\subsubsection{Cosmic-ray electron and positron background flux}
Throughout, we use an empirical parametrization of the local interstellar spectra (LIS) for $e^-$ and $e^+$ from Ref.~\cite{Bisschoff:2019lne}. This LIS fit is a data-driven description of the charged-particle flux outside the heliosphere; it is commonly treated as an ``astrophysical background'' in the sense that it represents the measured baseline spectrum against which additional components (such as DM injection) are constrained \cite{Bisschoff:2019lne,John:2021ugy}. In particular, the LIS fit implicitly captures the net contribution of primary sources (e.g. supernova remnants and pulsars) and secondary production, without assuming a specific source decomposition.
Ref.~\cite{Bisschoff:2019lne} calibrates the fit up to $\sim 100~\mathrm{GeV}$. Since our solar-halo dataset extends to $100~\mathrm{GeV}$ and since the IC kernel mixes a range of electron energies into each $\gamma$-ray bin, we extrapolate the LIS parametrization above $100~\mathrm{GeV}$ rather than imposing a hard cutoff. We consider this to be more physical than imposing an abrupt reduction of background contribution to zero, and expand on this quantitatively in Appendix~\ref{sec:Systematics}.
Namely, we use the following best-fit with $E_0 = \SI{1}{GeV}$
\begin{widetext}
\begin{align}
    J_{e^{-}} &= 255.0 \left(\frac{E}{E_0}\right)^{-1} \left(\frac{E/E_0 + 0.63}{1.63}\right)^{-2.43} + 6.4 \left( \frac{E}{E_0} \right)^2 \left( \frac{E / E_0 + 15.0}{16.0} \right)^{-26.0} \; , \label{eq:electron-background}\\
    J_{e^{+}} &= 25.0 \left(\frac{E}{E_0}\right)^{0.1} \left( \frac{(E/E_0)^{1.1} + 0.2^{1.1}}{1 + 0.2^{1.1}} \right)^{-3.31} + 23.0 \left(\frac{E}{E_0}\right)^{0.5} \left( \frac{E/E_0 + 2.2}{3.2} \right)^{-9.5} \; .\label{eq:positron-background}
\end{align}
\end{widetext}

\subsubsection{DM-induced flux}
DM decay injects relativistic $e^\pm$ throughout the Galactic halo with the differential source term
\begin{equation}
Q_{\chi}(E_{e},\mathbf{r})
  = \frac{\rho_{\chi}(\mathbf{r})}{m_{\chi}\tau_{\chi}}
    \,\frac{{\rm d}N_{e}}{{\rm d}E_{e}},
\label{eq:Qchi}
\end{equation}
where $\rho_{\chi}(\mathbf{r})$ is the DM density, $m_{\chi}$ its mass, $\tau_{\chi}$ the lifetime, and ${\rm d}N_{e}/{\rm d}E_{e}$ the $e^\pm$ spectrum per decay. In this work, we consider different decay channels, in each case making up for the entirety of the decay width.

Since the propagated local $e^\pm$ density from decaying DM is dominated by sources within a diffusion-loss length of the Solar position, the most important astrophysical input is the local DM density $\rho_{\rm DM, \odot} = 0.40\, \mathrm{GeV\,cm^{-3}}$~\cite{Cirelli:2010xx}, which we note is conservative relative to the most recent inferences by Refs.~\cite{Soding:2025mod,bienayme_dark_2024} suggesting central values in excess of $ 0.44\, \mathrm{GeV\,cm^{-3}}$ in the solar neighborhood. Changing the local DM density value simply rescales the constraints we derive linearly. In some DM scenarios, the solar gravitational potential can gravitationally focus the unbound halo DM distribution, mildly enhancing $\rho_\chi(r)$ toward small heliocentric radii. For an approximately Maxwellian halo, this enhancement is expected to be modest at Earth position and larger near the solar surface. Such a density enhancement is model dependent and would translate into a correspondingly larger ICS $\gamma$-ray signal, strengthening our constraints. A dedicated treatment of gravitational focusing, including the detailed escape/transport of decay products, is beyond the scope of this work.

The differential flux of electrons and positrons obtained at any point in our Galaxy for a source term $Q_{\chi}$ can then be written as
\begin{equation}
\frac{\dd \Phi_{e^\pm}}{\dd \Eein} (E_{e},\mathbf{r})
   = \frac{1}{4\pi b(E_{e})}
     \!\int_{E_{e}}^{m_{\chi}/2}
      {\rm d}E'\,Q_{\chi}(E',r) I(\Eein, E^\prime, \mathbf{r}),
\label{eq:Nchi}
\end{equation}
where $b(E_{e})$ is the total energy-loss rate (in ${\rm GeV\,s^{-1}}$) and $I$ is the generalized halo function \cite{Cirelli:2010xx}.

We obtain the propagated local $e^\pm$ spectra using \texttt{PPPC4DMID}~\cite{Cirelli:2010xx}, which implements the Green's function solution of the diffusion-loss equation for standard MIN/MED/MAX-like propagation setups.
Consequently, varying the Galactic DM density profile among standard choices (e.g.\@ NFW vs.\@ Einasto) produces only subdominant changes in the local injected density, while the choice of propagation model can produce a much more pronounced effect.
We explicitly quantify this dependence in Sec.~\ref{sec:Results}.

\subsubsection{Solar modulation}
Charged particles entering the heliosphere experience velocity changes by the heliospheric magnetic field. This effect of solar modulation of the interstellar charged particle fluxes in the heliosphere is an important aspect for a precise understanding of the production of solar ICS $\gamma$-rays. Following the force-field approximation~\cite{Gleeson:1968zza}, we model solar modulation via an effective solar modulation potential $\Phi(r)$, a function of the distance from the Sun, that maps the LIS spectrum to the heliospheric spectrum.
See Ref.~\cite{Linden:2025xom} for an explicit heliocentric IC implementation and discussion of the energy/angle dependence of modulation effects.
In the force-field approximation, the modulated differential intensity at radius $r$ is related to the unmodulated interstellar spectrum $J_{e^{\pm}}^{\rm LIS}$ from both the astrophysical background (Eqs.~\eqref{eq:electron-background}--\eqref{eq:positron-background}) and the DM-induced flux (Eq.~\eqref{eq:Nchi}),
\begin{equation}
    J_{e^\pm}(r,\Eein)
    = J_{e^{\pm}}^{\rm LIS}(\Eein) \frac{\Eein^2 - m_e^2}{(\Eein + e \Phi(r))^2 - m_e^2} \; .
    \label{eq:forcefield}
\end{equation}
The effect of solar modulation on the solar $\gamma$-ray halo has recently been studied in detail in Ref.~\cite{Linden:2025xom}. Following their findings, we adopt the spherically symmetric solar modulation potential
\begin{equation}
    \Phi(r) = \Phi_0 \frac{(r/\SI{1}{AU})^{-0.1} - (r_b/\SI{1}{AU})^{-0.1}}{1 - (r_b / \SI{1}{AU})^{-0.1}} \; ,
\label{eq:Phi_r}
\end{equation}
where $\Phi_0$ is the solar modulation potential evaluated on the Earth orbit and $r_{b}=\SI{100}{AU}$. 
In our analysis we profile over the solar modulation potential $\Phi_0$, and do so independently for the electron and positron fluxes to account for charge-dependent effects of the solar modulation. The effect of solar modulation is negligible at high energies corresponding to $\gamma$-ray energies above $\SI{3}{GeV}$.

The modulation potential varies over the solar cycle; the {\it Fermi}-LAT solar-halo dataset averages over many years, so our $\Phi_0^{e^\pm}$ should be interpreted as an effective, time-averaged modulation parameter consistent with the treatment in Ref.~\cite{Linden:2025xom}.

\subsection{Compton scattering cross section}\label{sec:compton}
\begin{figure}
    \centering
    \includegraphics[scale=1]{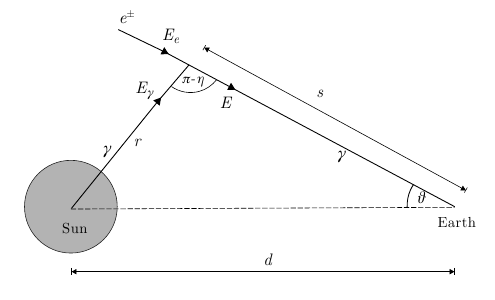}
    \caption{Geometry and kinematics for solar ICS in the SRF.}
    \label{fig:geometry}
\end{figure}
The Klein--Nishina formula describing the leading order cross section for photon-electron scattering in the ERF (referred to by primed quantities) reads
\begin{multline}
    \frac{\dd \sigma}{\dd \EpoutERF \dd \Omega^\prime} = \frac{r_e^2}{2} \left(\frac{\EpoutERF}{\EpinERF}\right)^2 \left( \frac{\EpoutERF}{\EpinERF} + \frac{\EpinERF}{\EpoutERF} - \sin^2\eta^\prime \right) \\
    \times \delta \left(\EpoutERF - \frac{\EpinERF}{(1 + \EpinERF (1 - \cos\eta^\prime) ) }\right),
    \label{eq:differentialcrossection}
\end{multline}
where $\Omega^\prime$ is the emission angle of the outgoing photon, and  $\eta^\prime$ is the photon scattering in the ERF, defined as $\cos\eta^\prime = \mathbf{p_{\gamma}^\prime} \cdot \mathbf{p^\prime} / (\EpinERF \EpoutERF)$. Here, $\EpinERF, \EpoutERF$ are the energies of the incoming and outgoing photon in the ERF respectively, and are related to the (unprimed) quantities in the solar rest frame (SRF) via
\begin{align}
    \EpinERF &= \frac{\Epin \Eein}{m_e} (1 - \cos\theta_{\rm in}) = \frac{\Epin \Eein}{m_e} (1 - \cos\eta) \; , \\
    \EpoutERF &=  \frac{\Epout m_e}{\Eein (1 + \cos\theta_{\rm out}^\prime)}
    = \frac{\Epout m_e}{\Eein (1 - \cos\eta^\prime)} \; .
\end{align}
$\theta_{\rm in/out}$ is the angle between the electron velocity and the incoming/outgoing photon in the SRF.
The above relations are valid in the ultra-relativistic limit $\beta \approx 1$, and where we used the fact that $\pi - \eta^\prime = \theta^\prime_{\rm out}$, see also Fig.~\ref{fig:geometry}.

We can now express the differential scattering cross section in the SRF in terms of the Klein--Nishina formula in the ERF using the variable transformations 
\begin{equation}
    \frac{\dd \EpoutERF}{\EpoutERF} = \frac{\dd \Epout}{\Epout} \; ,
    \quad
    \dd \Omega^\prime = \frac{\dd \Omega}{\Eein^2 (1 - \cos\eta^\prime)^2} \; .
\end{equation}
Hence the LHS of Eq.~\eqref{eq:differentialcrossection} becomes
\begin{equation}
    \frac{\dd \sigma}{\dd \Epout \dd \Omega} = 
    \frac{\Eein^2}{m_e^2} (1-\cos\eta^\prime)^2 \frac{\EpinERF}{\Epin} \frac{\dd \sigma}{\dd \EpinERF \dd \Omega^\prime} \; .
\end{equation}
Using this expression and evaluating the angular integral for the outgoing photon in Eq.~\eqref{eq:emissivity-simplified}, where we assume the cosmic-ray $e^\pm$ flux to be isotropic,  we arrive at
\begin{multline}
    j_\omega(\Epout,\mathbf{r},\eta)
    = \\
    \int \dd \Eein \dd \Epin \;  \frac{\dd \Phi_e}{\dd \Eein}(\Eein)  \frac{\dd N_\gamma}{\dd \Epin}(\Epin,\mathbf{r}) \frac{r_e^2}{2} \frac{1}{\Epin (\Eein - \Epout)^2} \\
    \times  \left[
        2 - 2 \epsilon \left(2 + \frac{m_e}{{\EpinERF}}\right) + \epsilon^2 \left( 3 + \frac{2 m_e}{{\EpinERF}} + \frac{m_e^2}{{\EpinERF}^2} \right) - \epsilon^3
    \right] \; ,
\label{eq:emissivity}
\end{multline}
where $\epsilon\!=\!\Epout/\Eein$.
In the above we treat the Sun as a point-source with regards to the directional information of the incident photon for the scattering angle but use the solar photon field density derived from the emission of a finite disk as in Eq.~\eqref{eq:solar-photon-density-integrated}. We comment on the error associated with the unidirectional approximation in Appendix~\ref{sec:Systematics}.

Further, the kinematics also imply
\begin{equation}
    \Epout \leq \frac{2 \Eein \EpinERF}{m_e + 2\EpinERF} \; .
\end{equation}
This upper bound on the outgoing photon energy $\Epout$ translates into a minimum injected electron or positron energy $\Eein$ as a function of the former
\begin{equation}
    \Eein \geq \frac{\Epout}{2} + m_e \sqrt{ \left(\frac{\Epout}{2 m_e} \right)^2 + \frac{\Epout}{2 \Epin (1-\cos\eta)} } \; .
\end{equation}

\subsection{Line-of-sight intensity and observed \texorpdfstring{$\gamma$}{γ}-ray flux}\label{sec:los}
The observable intensity per steradian is obtained by integrating the emissivity along the line-of-sight,
\begin{equation}
I(\Epout,\theta) 
 = \int\!\dd s\,
   j_\omega(\Epout,\mathbf{r}(s,\theta),\eta(s,\theta)) \; ,
\label{eq:intensity}
\end{equation}
where $\theta$ is the angular distance from the solar center and $s$ the line-of-sight coordinate. Here, $r(s,\theta) = \sqrt{d_\odot^2 + s^2 - 2 s d_\odot \cos\theta}$ and $\eta$ is determined by the geometry, see Fig.~\ref{fig:geometry}. In the ultra-relativistic limit, the scattered $\gamma$-ray direction is well approximated as collinear with the incident $e^\pm$, so $\eta$ is determined purely by solar geometry.

The predicted flux spectrum is then
\begin{equation}
\frac{{\rm d}\Phi}{\dd\Epout}
 = \frac{1}{\Omega_{\rm ROI}} \int_{\Omega_{\rm ROI}} I(\Epout,\theta)\,{\rm d}\Omega \; ,
 \label{eq:flux}
\end{equation}
evaluated over the region of interest $\Omega_{\rm ROI}$, which we choose as annuli around the Sun.

Eqs.~(\ref{eq:Qchi})–(\ref{eq:flux}) together outline the full heliocentric ICS calculation.

\section{Analysis}\label{sec:Analysis}
In this section we describe the dataset and statistical procedure used to derive constraints on the DM lifetime.
\subsection{Data}\label{sec:data}
We use the binned solar-halo $\gamma$-ray flux measurements derived from 15 years of \textit{Fermi}-LAT observations in Ref.~\cite{Linden:2025xom}. That work develops a dedicated analysis pipeline for moving sources, converting both photon counts and exposure into helioprojective coordinates centered on the instantaneous solar position and building a data-driven background model from sky regions observed when the Sun is elsewhere on the sky. This procedure enables a high-significance measurement of the solar ICS halo while controlling the dominant backgrounds and instrument-specific systematics associated with the Sun's unique location in the \textit{Fermi}-LAT instrumental coordinate system (e.g. non-uniform sampling in the azimuthal coordinate $\phi$ due to the spacecraft solar-panel pointing constraint).

The dataset is provided in 28 logarithmically-spaced energy bins spanning $E \in [31.6~\mathrm{MeV},\,100~\mathrm{GeV}]$, and in a set of concentric annuli about the solar center extending to $\theta=45^\circ$. The disk emission in the angular region $\theta<0.26^\circ$ is removed in the published data.

Ref.~\cite{Linden:2025xom} shows that the solar halo is robustly detected in essentially all energy bins and annuli, but that the analysis is systematics dominated in specific regimes. The non-solar diffuse background is much smaller than the halo surface brightness.

Our goal is not to reanalyse the solar halo, but instead to leverage the measured halo fluxes as an indirect detection dataset for decaying DM. We therefore keep the experimental extraction fixed (as given in Ref.~\cite{Linden:2025xom}) and propagate DM-injected $e^\pm$ through the solar ICS calculation to construct signal templates in the same binning for a direct statistical comparison.

\subsection{Background and bias}\label{sec:bkg-and-bias}
The reliability of the limits derived from the profile likelihood ratio depends on the goodness-of-fit of the null hypothesis (background-only model).
In our analysis, we find that while the high-energy $\gamma$-ray bins above \SI{3}{GeV} exhibit a good fit with a $\chi^2$ per degree-of-freedom of $\chi_\nu^2 \approx 1.25$, the low-energy regime (\SI{31.6}{MeV} to \SI{3}{GeV}) suffers from a significant background bias, with $\chi_\nu^2 \approx 4.5$. 
This discrepancy is primarily attributed to the complexities of modeling the solar modulation potential $\Phi(r)$, which governs the cosmic-ray $e^\pm$ spectra in the inner heliosphere~\cite{Linden:2025xom}.

Rather than relying on an imperfect physical model for the solar modulation at low energies, a more complete physical model of heliospheric transport could reduce residuals, but developing such a model is beyond the scope of this work. We therefore assume there exists an unknown systematic uncertainty associated with the background modeling that dominates the low-energy bins. 
To account for this, we introduce a relative systematic uncertainty $\sigma_{\text{sys}}$ which is added in quadrature to the statistical uncertainties $\sigma_{ij}$ in the likelihood function. The total uncertainty for each bin is then modeled as
\begin{equation}\label{eq:totaluncertainty}
    \sigma_{ij, \text{tot}}^2 = \sigma_{ij}^2 + {(\sigma^{\rm sys}_{ij})}^2 = \sigma_{ij}^2 + (\epsilon_{\rm sys} \cdot b_{ij}(\boldsymbol{\theta}))^2 \; ,
\end{equation}
where $b_{ij}$ is the binned background model flux and $\epsilon_{\rm sys}$ is the fractional systematic error we assume for energy bins $i=1, \dots, 16$ corresponding to $E<\SI{3}{GeV}$. The use of a relative uncertainty is physically motivated by the fact that the solar $\gamma$-ray flux varies by several orders of magnitude across different annuli due to the quadratic fall-off with distance from the Sun of the solar photon field and the suppression of the cross section and the resulting $\gamma$-ray spectrum at high energies. 
A constant (additive) systematic error would fail to capture the background variance across the large dynamic range of our region of interest. 
By inflating the variance in this manner, we ensure that the resulting limits on the DM lifetime $\tau_\chi$ are conservative and account for the current theoretical limitations in modeling low-energy solar IC emission. We determine $\epsilon_{\rm sys}$ by requiring that the background-only model yields $\chi^2_\nu \simeq 1$ in the low-energy subset of bins, thereby treating the observed variance excess as an effective, energy-independent fractional modeling uncertainty in that regime. We stress that this choice ensures our limits are robust: increasing $\epsilon_{\rm sys}$ can only weaken the constraint on $\tau_\chi$.

\begin{figure}
    \centering
    \includegraphics[scale=1.0]{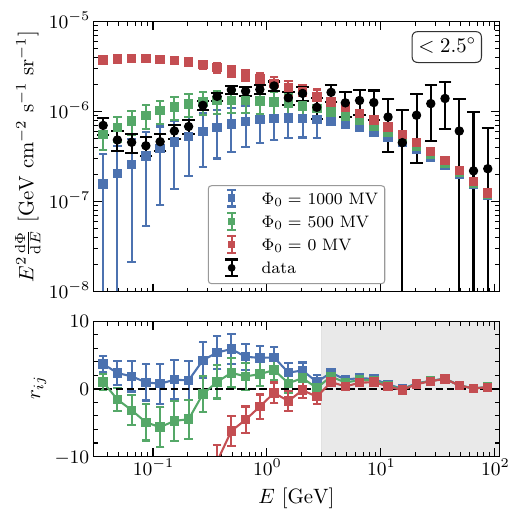}
    \caption{$\gamma$-ray flux and IC photon flux from background electron and positron fluxes. {\bf Top:} Measured gamma ray flux (black) compared to the background IC $\gamma$-ray flux (blue, green, red) with an assumed constant relative systematic uncertainty of $\sim 30\%$.
    {\bf Bottom:} Residuals between the data and the background flux for different values of the solar modulation potential, including a systematic uncertainty on the background shown as error bars. The IC photon flux is largely unaffected by solar modulation at high energies (gray-shaded).}
    \label{fig:bkg-data-residuals}
\end{figure}

In Fig.~\ref{fig:bkg-data-residuals} we show the $\gamma$-ray flux resulting from the interstellar background $e^{\pm}$ spectrum for different values of the solar modulation potential for the innermost annulus around the Sun. We also show the residuals $r_{ij} = ( d_{ij} - b_{ij}(\boldsymbol{\theta}))/\sigma_{ij} \pm (\sigma^{\rm sys}_{ij} / \sigma_{ij})$.
Here we choose the same values for the solar modulation potentials of electron and positron for the purpose of illustration.

This behavior is consistent with the systematics discussion in Ref.~\cite{Linden:2025xom}. 
Rather than over-interpreting structured residuals in the low-energy regime where solar modulation dominates, we absorb the excess variance into the effective fractional uncertainty in Eq.~\eqref{eq:totaluncertainty}. Moreover, the Sun occupies a special region of the \textit{Fermi}-LAT instrumental coordinate system (owing to the solar-panel pointing constraint), so residual $\phi$-dependent effective-area and cosmic-ray leakage effects can imprint energy-dependent artifacts that are difficult to diagnose with standard blank-sky tests~\cite{Linden:2025xom}. Our strongest constraints are driven by higher-energy bins where the background-only fit quality is good and solar modulation effects are subdominant.

\subsection{Signal templates}\label{sec:signal}
The DM signal component $s_{ij}(\tau_\chi,m_\chi, \boldsymbol{\theta})$ is constructed by generating templates for the expected $\gamma$-ray flux across the defined energy bins and solar annuli. We compute the templates for DM masses in the range $m_\chi \in [\SI{10}{GeV},\SI{10}{TeV}]$ in various decay channels, such as $\chi \to e^{+}e^{-}$.
Since the signal flux scales linearly with the decay rate $\tau_\chi^{-1}$, we only compute the templates for one reference lifetime of $\tau_\chi\!=\!\SI{e25}{s}$.

The DM-induced signal extends far away from the Sun similar to the background flux, up to our maximum elongation angle of $45^\circ$ from the Sun, and energies up to around half the DM mass. We find the following:
\begin{itemize}
    \item {\bf Angular morphology.}
    The intensity peaks towards the Sun and falls off with angular distance from the Sun as $\theta^{-1}$, similar to the background. This is because the morphology is primarily determined by the density of the solar photon field, which drops quickly with distance from the Sun as  $r^{-2}$. The LOS integral~\eqref{eq:intensity} is therefore typically dominated by spatial contributions closest to the Sun. At very large elongation angles, where the solar photon density is rather constant along the line-of-sight, the intensity is also controlled by the geometry and spectral shape of the $e^\pm$ flux due to the anisotropy of the scattering kernel.
    
    \item {\bf Spectral shape.} The $\gamma$-ray spectrum inherits a sharp high-energy cutoff set by the injected prompt cosmic ray $e^{\pm}$ spectrum in the $\chi\to e^{+}e^{-}$ channel ($E_e \lesssim m_\chi/2$ for two-body decays) together with Klein--Nishina suppression at large $E_e E_\gamma$. The spectrum also features a long low-energy tail resulting from the energy-spread propagated cosmic-ray $e^\pm$ spectrum and the enhanced cross section at lower energies.
    Solar modulation affects the low-energy tail primarily below a few GeV in $\gamma$-ray energy.
\end{itemize}
We further describe the qualitative and quantitative features of the signal spectrum, which is very distinct from the background and background-induced $\gamma$-ray spectrum below.
The position of the signal peak can be approximately described by a power-law fit as a function of the DM mass for the $e^{\pm}$ decay channel with
\begin{equation}
    E_{\mathrm{peak}}(m_\chi)=\frac{\frac{4}{3}\gamma_e^2\,\epsilon_{\mathrm{ph}}}{1+4\gamma_e\epsilon_{\mathrm{ph}}/m_e},\qquad \gamma_e=\frac{m_\chi}{2m_e},
    \label{eq:peak-energy}
\end{equation}
where $\gamma_e$ is the Lorentz factor for the kinematic endpoint of the decay into $e^\pm$ and $\epsilon_{\rm ph} \approx \SI{0.4}{eV}$ is a typical solar photon energy, The signal scales linearly with the DM decay rate. The spectrum has a long low-energy tail and drops off rapidly above the peak, and the amplitude is only  mildly dependent on the DM mass. We provide detailed fits for the unmodulated signal spectrum in Appendix~\ref{sec:channels}.
\begin{figure}
    \centering
    \includegraphics[scale=1.0]{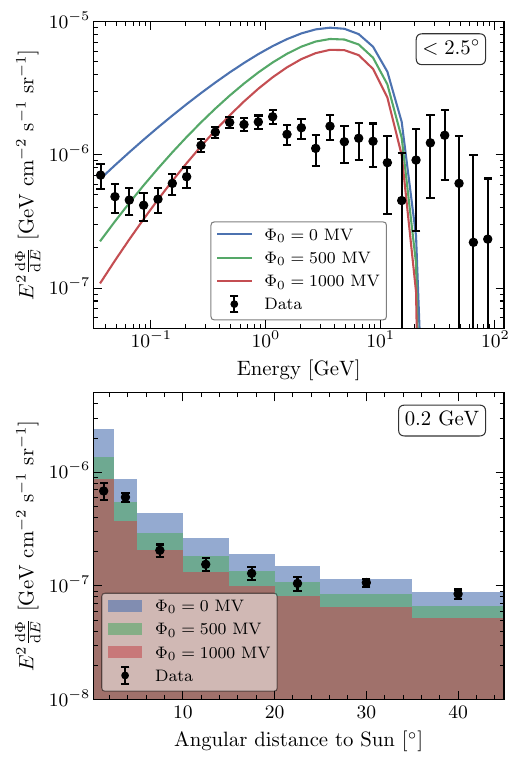}
    \caption{Example of the DM-induced flux from decay via $\chi \to e^+ e^-$ compared to the data for a benchmark template with $m_{\chi}=\SI{100}{GeV}$ and $\tau_\chi=\SI{e25}{s}$. \textbf{Top:} Signal spectrum for the innermost annulus within $0.26^\circ < \theta < 2.5^\circ$ of the Sun. \textbf{Bottom:} Signal morphology in the energy bin centered at $\SI{0.2}{GeV}$.}
    \label{fig:signal-morphology}
\end{figure}

\subsection{Likelihood}\label{sec:likelihood}
We use the flux estimators with approximately Gaussian errors after the full LAT event-level analysis (exposure, effective area, and background subtraction) and therefore we adopt a Gaussian likelihood rather than a Poisson likelihood for raw counts. We additionally treat different energy bins and annuli as independent, i.e.\@ we use a diagonal covariance approximation; potential correlations are subdominant for our constraints because the signal has a distinct joint energy--angle
dependence.

We model flux data $d_{ij}$ per energy bin $i$ and annulus $j$ by a background component $b_{ij}(\boldsymbol{\theta})$ and DM-induced signal component $s_{ij}(\tau_\chi,m_\chi,\boldsymbol{\theta})$ as a function of the model parameters as described in Sec.~\ref{sec:Formalism}.
Treating the data across energy bins and annuli as statistically independent, we build the log-likelihood function
\begin{equation}
    \ell(\tau_\chi, m_\chi, \boldsymbol{\theta})=-\frac{1}{2}\sum_{i,j}
        \frac{ (d_{ij} - s_{ij}(\tau_\chi, m_\chi, \boldsymbol{\theta}) - b_{ij}(\boldsymbol{\theta}) )^2 }{\sigma_{ij,\rm tot}^2},
        \label{eq:loglikelihood}
\end{equation}
where $\sigma_{ij,\rm tot}$ is the total uncertainty as given in Eq.~\eqref{eq:totaluncertainty}. The model components $s_{ij}, b_{ij}$ are calculated from flux templates for the bins for a given DM mass. More specifically, $s_{ij}(\tau_\chi, m_\chi, \boldsymbol{\theta}) = \mathcal{A}(\tau_\chi) T_{ij}^{\rm sig}(m_\chi, \boldsymbol{\theta})$, where $\mathcal{A}(\tau_\chi)$ is the normalization of the signal flux scaling linearly with the DM lifetime $\tau_\chi$ and $b_{ij}(\boldsymbol{\theta}) = T_{ij}^{\rm bkg}(\boldsymbol{\theta})$. The signal and background templates are computed for a discrete set of the shared nuisance parameters $\boldsymbol{\theta}$, the solar modulation potential normalizations for electron and positrons, $\Phi_0^{e^\pm}$.

The profile log-likelihood as a function of the parameter of interest $\mu$ (the DM lifetime $\tau_\chi$) is given by maximizing the log-likelihood function over the nuisance parameters~$\boldsymbol{\theta}$,
\begin{equation}
    \ell_p(\mu) =  \max_{\boldsymbol{\theta}} \ell(\mu,\boldsymbol{\theta}).
\end{equation}
Let us also define the maximum likelihood estimator (MLE)
\begin{align}
    (\hat{\mu},\boldsymbol{\hat{\theta}}) &= \arg\max_{\mu,\boldsymbol{\theta}} \ell(\mu,\theta) \; .
\end{align}

In order to determine an upper bound for the parameter of interest $\mu$, we define the following test statistic,
\begin{equation}
    q_\mu = 
    \begin{cases}
        -2\left( \ell_p(\mu)-\ell_{\rm max}\right), & \hat{\mu}\leq \mu \, , \\
        0, & \hat{\mu} > \mu \, ,
    \end{cases}
    \label{eq:test-statistic}
\end{equation}
where $\ell_{\rm max} = \ell(\hat{\mu},\boldsymbol{\hat{\theta}})$, and which is based on the logarithm of the profile likelihood ratio, $-2 (\ell_{p}(\mu) - \ell_{\rm max})$. The reason for setting $q_{\mu} = 0 $ for $\hat{\mu} > \mu$ is that when setting an upper limit, one would not regard data with $\hat{\mu} > \mu$ as representing less compatibility with $\mu$ than the data obtained, and therefore this is not taken as part of the rejection region of the test. From the definition of the test statistic one sees that higher values of $q_{\mu}$ represent greater incompatibility between the data and the hypothesized value of the parameter of interest $\mu$ \cite{Cowan:2010js}.

\subsection{Upper limits}\label{sec:upper-limits}
Using the test statistic defined in Eq.~\eqref{eq:test-statistic}, we set an upper limit on the DM lifetime using the following methodology.
\paragraph{Upper limits at 95\% confidence level}
The upper limit on the parameter of interest, the DM lifetime, at 95\% confidence level (CL) corresponds to $\mu_{95}$ with
\begin{equation}
    \mathrm{Prob}(q_\mu > q_{\mu_{95}})=0.05.
\end{equation}
After Wilks' theorem, our test statistics can be approximated by a $\chi^2$-distribution with one degree of freedom, such that the one-sided 95\% CL\@ upper limit can be set by
\begin{equation}
    q_\mu < \chi^2_{1;0.90} = 2.71 .
\end{equation}

\paragraph{Expected upper limits}
For the expected limit we use an Asimov dataset constructed from the best-fit background-only model, $d^{A}_{ij}=b_{ij}(\hat{\boldsymbol{\theta}}_{0})$, where $\hat{\boldsymbol{\theta}}_{0}$ maximizes the likelihood under the null hypothesis for the existing dataset and background model. We then compute the likelihood ratio on this Asimov dataset for the fixed nuisance parameters $\hat{\boldsymbol{\theta}}_{0}$ as it is the actual background template compared to the data that drives the determination of the nuisance parameters. We do this since the MLE for profiling only with the signal would be unbounded, always preferring a strong suppression of any additional signal when the data is fixed to the background and the signal is a strictly monotonically decreasing function of the solar modulation potential. This amounts to modifying the log-likelihood in Eq.~\eqref{eq:loglikelihood} to
\begin{equation}
    \ell_{A}(\tau_\chi, m_\chi)=-\frac{1}{2}\sum_{i,j}
        \frac{ (s_{ij}(\tau_\chi, m_\chi \hat{\boldsymbol{\theta}}_0))^2}{\sigma_{ij,\rm tot}^2}.
\end{equation}

\paragraph{Power-constrained limits}\label{sec:pcl}
It is more conservative to claim at most the expected upper limit from the Asimov dataset to avoid overly aggressive constraints resulting from a downward fluctuation in the data. Taking the minimum upper limit of the expected and the 95\% CL upper limits yields the power-constrained limit (PCL) \cite{Cowan:2010js}, which we report in this work.
The power-constrained upper limit corresponds to
\begin{equation}
    \tau_{\rm PCL} = \min \, \lbrace \tau_{q_{\mu}} , \tau_{\rm exp} \rbrace.
\end{equation}

\section{Results}\label{sec:Results}
\begin{figure*}[t]
    \includegraphics[scale=1.0]{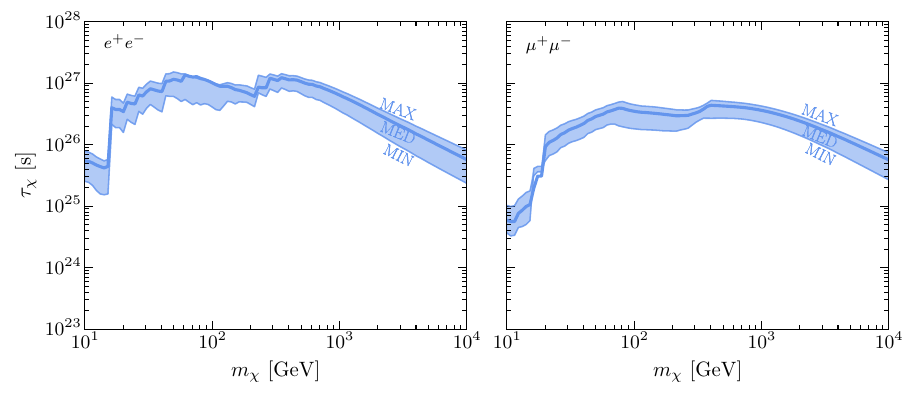}
    \caption{Upper limit on the lifetime $\tau_\chi$ of DM decaying into $e^+e^-$ (left panel) and $\mu^+\mu^-$ (right panel), based on $\gamma$-ray emission in the solar halo.}
    \label{fig:e-vs-mu}
\end{figure*}

Fig.~\ref{fig:e-vs-mu} shows our main result for the exclusion limits on the DM lifetime in the benchmark channel $\chi\to e^+e^-$ in the left panel, and the $\chi \to \mu^+ \mu^-$ channel in the right panel, representative of channels with softer injection spectra.

For DM with masses in the GeV to TeV range and decay into $e^+ e^-$, the solar $\gamma$-ray halo measurement excludes lifetimes up to $\tau_\chi \sim 10^{27}\,\mathrm{s}$, with the strongest sensitivity obtained when the upscattered photon spectrum falls within the \textit{Fermi}-LAT energy range where the measured solar-halo flux is statistically precise and solar modulation is subdominant (typically $E_\gamma \gtrsim$ a few GeV).
The constraining power arises primarily from the spectral shape of the signal: the spectrum exhibits a hard cutoff above $E_{\rm peak}$ (see Eq.~\eqref{eq:peak-energy}) that is absent in the astrophysical halo, which extends smoothly to high energies following the cosmic-ray $e^\pm$ spectrum of astrophysical origin. The morphology, with the intensity peaking toward the Sun as $\theta^{-1}$, is shared by both components, but serves to confirm the solar IC origin of any excess and to separate it from unrelated diffuse backgrounds.

The upper limit on the DM lifetime is of the order of $\SI{e26}{s}$ for low masses where the signal falls entirely within the low-energy bins dominated by solar modulation. The limits are strengthened by an order of magnitude above $\sim \SI{20}{GeV}$, where the cutoff energy enters the regime that is insensitive to the solar modulation uncertainties. For masses up to $\SI{500}{GeV}$, the lifetime is generally constrained at the level of $\tau_\chi \simeq \SI{e27}{s}$, and softens for larger masses due to the energy-suppression of the scattering cross section, and because the bulk of the signal spectrum falls outside of the range of observation of \textit{Fermi}-LAT. We note that since the signal is determined by the local DM density, the limits are robust to large-scale halo profile uncertainties. 
The band spanned by the MIN and MAX propagation models remains within a factor of two of the MED result across the entire mass range, demonstrating that the limits are robust to uncertainties in the cosmic-ray $e^\pm$ propagation model. We quote the MED limit as our nominal result.

Compared to $e^+e^-$, the signal spectrum for the $\mu^+ \mu^-$ channel is softened by the secondary decay kinematics, which reduces the peak energy compared to the sharp high-energy cutoff and typically weakens the limits at fixed mass. Nevertheless, we find that solar ICS remains a strong probe for $\mu^+\mu^-$ over a wide mass range, with limits within a factor of a few of the $e^{+}e^{-}$ channel: the lifetime for the $\mu^{+}\mu^{-}$ decay channel is constrained up to the level of $\sim \SI{e25}{s}$ for low masses around $\SI{10}{GeV}$, and stronger constraints of $\text{few} \times \SI{e25}{s}$ are again set for masses above $\SI{20}{GeV}$ where the signal lies within the \textit{Fermi}-LAT energy window. Because of the softer injection spectrum, the constraints remain stable up to TeV masses since the bulk of the signal stays within the \textit{Fermi}-LAT energy range to higher masses than for $e^+e^-$. We provide results for additional channels in Appendix~\ref{sec:channels}. We find that, with even heavier primary decay products and thus even softer secondary $e^\pm$ spectra, the constraints are weakened particularly at low DM mass.

\begin{figure}[t]
    \centering
    \includegraphics[scale=1.0]{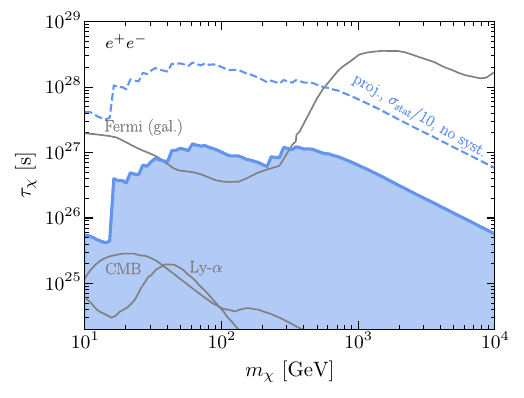}
    \caption{Comparison of the most recent $\gamma$-ray and cosmological constraints (gray) on decaying DM in the $e^{+}e^{-}$ channel with the PCL limits at 95~\% CL\ from the solar $\gamma$-ray halo measurement (blue-shaded). Prospective limits on decaying DM from solar $\gamma$-ray halo measurements are indicated by the dashed blue line.}
    \label{fig:e-gammaray-limits}
\end{figure}

In Fig.~\ref{fig:e-gammaray-limits} we compare our solar ICS limits to recent $\gamma$-ray constraints on decaying DM in the $e^+e^-$ channel. Solar ICS bounds provide an independent cross-check of the allowed lifetime parameter space, especially for leptonic channels where ICS is the dominant $\gamma$-ray production mechanism. We also show prospects of the solar halo as a probe of decaying DM in Fig.~\ref{fig:e-gammaray-limits} (dashed blue), where we assume an idealized dataset without systematics and a statistical uncertainty reduced by an order of magnitude representing an optimistic projection. It should be noted that the master dataset from which the fluxes were extracted for this work had significant cuts applied to the number of events already. These removed around 90\% of the events to obtain the low-background dataset used in this work. Accounting for potential improvements in exposure, background modeling, and angular resolution with future solar $\gamma$-ray observation, we consider this an instructive baseline projection. We show that the constraints are strengthened significantly across the entire mass range $\SI{10}{GeV}$ to $\SI{10}{TeV}$ by typically just more than one order of magnitude, but closely following the overall shape of our limits established in this work.

As a benchmark, we compare our constraints against decaying DM limits from complementary probes for the $e^+e^-$ channel. DM decay can impact the ionization and thermal history of the universe. Ref.~\cite{Liu:2016cnk} computed the impact of DM decay on cosmic reionization, via the production for electrons, positrons and photons. The possible heating and ionization impact of DM decay products results in a robust bound on lifetimes of $\sim \SI{e25}{s}$ between DM masses of $\SIrange{10}{100}{GeV}$ for the $e^+e^-$ channel. Ref.~\cite{Liu:2020wqz} computed energy injection heating bounds from measurements of the Lyman-$\alpha$ forest accounting for both photoionization and photoheating processes. These constraints are insensitive to the uncertainties of reionization and complement the CMB constraints. They also bound DM lifetime to around $\sim \SI{e25}{s}$ over a narrower DM mass range of $\SIrange{20}{80}{GeV}$, losing strength at higher mass. 

A non-cosmological search of interest for comparison are Galactic $\gamma$-ray constraints on decaying DM. Ref.~\cite{Cohen:2016uyg} utilized the \textit{Fermi}-LAT measurement of the diffuse Galactic $\gamma$-ray flux, and derived some of the most stringent constraints on DM lifetimes across the mass range from hundreds of MeV to above an EeV. This analysis included contributions to the DM-induced $\gamma$-ray flux resulting from both primary emission and IC scattering of primary electrons and positrons. Our constraint outperforms the Galactic $\gamma$-ray constraints from around $\SIrange{10}{300}{GeV}$ with the nominal MED propagation setup. 

Direct positron constraints from AMS-02, such as Ref.~\cite{Nguyen:2024kwy}, are among the strongest for DM direct decay into $e^+e^-$, with lifetimes up to $\SI{e25}{s}$ for DM masses below $\SI{10}{GeV}$ and up to $\SI{e29}{s}$ for masses above $\SI{10}{GeV}$ being excluded respectively. However, these lose strength when the final states are heavier or for DM masses above a TeV since AMS-02 has a sharp energy cutoff at these energies. Our constraints extend to larger DM masses due to \textit{Fermi}-LAT having larger energy reach in $\gamma$-rays. Recently, Ref.~\cite{Hiroshima:2025jyz} performed a combined analysis including CALET, DAMPE, H.E.S.S., positron flux from AMS-02, and $\gamma$-ray flux from HAWC, GRAPES-3 and CASA-MIA. Their analysis also found that AMS-02 gives the strongest limit on the DM lifetime, excluding $\tau_\chi \lesssim 10^{27}$ s for their minimum DM masses of around a TeV, which is at a comparable strength to our result. Extragalactic searches of decaying DM such as Ref.~\cite{Blanco:2018esa} leverage the IC process whilst accounting for the full evolution of cosmological electromagnetic cascade. These yield similarly strong lifetime constraints, at the $10^{27}$--$10^{28}$ s level, providing another valuable cross-examination of the relevant parameter space. 

If we consider heavier channels such as $\mu^+\mu^-$, our constraint outperforms Ref.~\cite{Cohen:2016uyg} over an intermediate DM mass range of $\sim\SIrange{50}{400}{GeV}$, and is also stronger than cosmological energy-injection bounds from the CMB and Lyman-$\alpha$ forest over the mass range shown. We find that our limits are typically somewhat weaker than reported direct positron constraints (e.g.\ Ref.~\cite{Ibarra:2013zia}), but, as in the $\chi\to e^+e^-$ case, they extend robustly to higher masses because the signal is mapped into $\gamma$-rays within the \textit{Fermi}-LAT energy window.

The physical reason for this complementarity is that for heavier channels direct positron searches lose a key advantage present in $\chi\to e^+e^-$: the line-like injection at $E\simeq m_\chi/2$ is replaced by a broad continuum after decay kinematics, showering, and propagation, reducing discrimination against smooth astrophysical backgrounds. By contrast, the solar-ICS signal is anchored to the solar photon field and retains a characteristic spectral turnover and high-energy cutoff set by the highest-energy $e^\pm$ in the decay cascade, with Klein--Nishina suppression at the highest energies. As a result, solar-ICS can be competitive with, and in parts of parameter space exceed, conventional Galactic diffuse $\gamma$-ray bounds in the regime where (i) the cutoff lies within the LAT energy range and (ii) the local ``solar converter'' enhancement offsets the smaller DM column.

At higher masses the situation reverses: diffuse Galactic $\gamma$-ray searches strengthen because they integrate ICS emission over kpc-scale DM columns and access higher-energy photons where backgrounds are reduced, while our solar-ICS analysis becomes limited by the LAT energy reach and the local nature of the target. In this sense solar-ICS provides a complementary \emph{local} $\gamma$-ray probe that bridges direct charged-particle searches (most powerful for hard leptonic channels at low-to-intermediate masses) and Galactic diffuse $\gamma$-ray analyses (most powerful at higher masses and for channels with substantial electromagnetic yield).

Finally, we note that our predicted astrophysical solar ICS component, using the LIS $e^\pm$ flux model of Ref.~\cite{Bisschoff:2019lne} and the modulation prescription of Ref.~\cite{Linden:2025xom}, reproduces the overall normalization and angular dependence of the measured halo within $\sim 10$\%, supporting the robustness of the constraints presented here.
We discuss further validation checks in Appendix~\ref{sec:Systematics}.

\section{Discussion and conclusion}
\label{sec:Conclusion}
We have shown that the measured solar ICS $\gamma$-ray halo can be repurposed as a dedicated indirect detection dataset for decaying DM. The essential mechanism is local: energetic $e^\pm$ injected by DM decay in the nearby halo traverse the inner heliosphere and upscatter the intense solar radiation field into $\gamma$-rays. This turns the Sun into an efficient, geometrically well-defined ``converter'' of leptonic energy injection into an extended, degree-scale $\gamma$-ray signal.

Using 15 years of \textit{Fermi}-LAT solar-halo measurements~\cite{Linden:2025xom}, we constructed energy- and annulus-binned signal templates in the full heliocentric geometry and derived conservative constraints while profiling over solar-modulation nuisance parameters. For leptonic decay channels we obtain competitive constraints, reaching lifetimes of order $\tau_\chi \sim 10^{27}\,\mathrm{s}$ in the GeV--TeV mass range. The constraining power is driven primarily by the unique DM signal structure relative to the astrophysical background component.

Our limits are systematically distinct from conventional Galactic diffuse $\gamma$-ray analyses and direct charged-particle searches. Galactic-halo $\gamma$-ray limits benefit from large DM columns whereas the solar ICS signal is anchored to a moving, local target with a sharply specified radiation field and a measured halo component. Direct $e^\pm$ measurements can be extremely powerful for hard leptonic final states, but are sensitive to local charged-particle systematics and have limited leverage at the highest energies set by instrumental reach; the solar ICS channel instead accesses the same underlying $e^\pm$ injection through a $\gamma$-ray observable and remains sensitive up to multi-TeV masses within the \textit{Fermi}-LAT energy window via the upscattered spectrum. In this sense, solar ICS provides a complementary local probe for the decaying-DM parameter space with different uncertainties.

The principal systematics in our analysis arise from (i) Galactic $e^\pm$ propagation modeling, which sets the local DM-induced $e^\pm$ spectrum and produces the dominant spread in the inferred lifetime limits, and (ii) heliospheric modulation at low energies, where current force-field descriptions are imperfect. We have treated the latter conservatively via variance inflation in the modulation-dominated regime and find that the resulting bounds are stable against reasonable variations of this procedure. Other effects, including the choice of Galactic DM density profile and approximations in the solar photon-field treatment, are subdominant at the level of precision relevant here. 

Looking ahead, the solar ICS channel is poised to improve. Higher-exposure measurements, improved control of low-energy systematics, and more realistic heliospheric transport models will directly sharpen the sensitivity of this method. Additionally, ICS may profer unique advantages at extremely high energies ($\gtrsim100$ TeV) due to the fast ICS energy loss time scales and larger instruments (compared to cosmic-ray experiments at TeV scales). However, such a search require experiments such as HAWC \cite{HAWC:2022khj} or LHAASO \cite{Zhou:2016ljf,LHAASO:2019qtb} to perform a dedicated solar halo analysis. More broadly, our results establish solar ICS as a viable and competitive component of the decaying DM search space, opening a local route to testing DM scenarios that inject energetic particles in the Galactic halo.

\acknowledgments
\section{Acknowledgments}
We would like to thank Tim Linden for providing the solar $\gamma$-ray flux data relevant for the constraints in this work.. We would like to thank both Tim Linden and Pedro De La Torre Luque for providing useful feedback on an earlier version of this manuscript. MD is supported by a KCL NMES Faculty Studentship. SB is supported by the Science and Technology Facilities Council (STFC) under grants ST/X000753/1 and ST/T00679X/1. 

\bibliographystyle{apsrev4-1}
\bibliography{references.bib}

\clearpage
\onecolumngrid
\appendix

\section{Systematics tests}\label{sec:Systematics}
We conduct a range of tests to assess the robustness of our constraints on the DM lifetime $\tau_\chi$ against modeling choices and systematic uncertainties. The dominant systematic uncertainty, namely the dependence on the assumed cosmic-ray propagation model, is discussed in the main text (Sec.~\ref{sec:Results}). Here we discuss the remaining, subdominant systematics and cross-checks, covering uncertainties in the DM-induced signal modeling, approximations in our flux template calculation, and robustness checks on our statistical treatment.

\subsection{Dark matter halo profile}
Our \textit{local} probe of DM decay is largely insensitive to the  halo profile: the relevant $e^\pm$ typically originate from within a diffusion-loss length of the Sun ($\lesssim \mathcal{O}(1)\,\mathrm{kpc}$), so the constraints depend primarily on the local DM density $\rho_\odot$ rather than the large-scale halo shape. We verify this explicitly by computing limits for NFW~\cite{Navarro:1995iw}, our baseline profile, alongside the isothermal~\cite{Bahcall:1980fb} and Einasto~\cite{Einasto:1965czb} profiles, each normalized to $\rho_\odot \! = \! 0.4\,\mathrm{GeV\,cm^{-3}}$~\cite{Cirelli:2010xx}. The resulting limits differ only at the percent level (which is also the level of precision of our numerical integration) across the entire mass range, confirming that the halo profile choice is negligible. Since the DM source term scales linearly with $\rho_\odot$ (Eq.~\eqref{eq:Qchi}), so do the lifetime limits, and the current uncertainty on $\rho_\odot$ of $\sim 10\%$, with some analyses preferring values as high as $0.44\,\mathrm{GeV\,cm^{-3}}$~\cite{Soding:2025mod,bienayme_dark_2024}, translates directly into a $\sim 10\%$ uncertainty on our constraints, which we regard as subdominant.

\subsection{Solar modulation potential}
The solar modulation potential varies significantly over the solar cycle and is not independently fixed by any external dataset in our analysis. We therefore profile over the modulation potentials $\Phi_{e^-}$ and $\Phi_{e^+}$ for electrons and positrons independently to account for charge-sign--dependent effects, treating them as nuisance parameters. The MLE for these parameters is driven primarily by the background flux templates, with the signal component shifting the MLE by $\mathcal{O}(\SI{50}{MV})$ depending on the template shape. For a background-only fit we find $(\Phi_{e^-}, \Phi_{e^+}) = (\SI{500}{MV}, \SI{100}{MV})$, in reasonable agreement with Ref.~\cite{Linden:2025xom}. We scan in steps of $\SI{25}{MV}$ over $\Phi_i \in [\SI{0}{GV}, \SI{1}{GV}]$, which is sufficiently fine since our MLEs always remain within this range. Solar modulation significantly affects the $e^\pm$ flux only below a kinetic energy corresponding to $E_\gamma \lesssim \SI{3}{GeV}$; at higher energies the effect is at the few-percent level. The overall impact on the PCLs is subdominant compared to the propagation uncertainty.

It is possible to use more detailed cosmic-ray propagation setups, see e.g.\@ \cite{Bisschoff:2019lne}, to determine the effect of solar modulation. However, as emphasized in Sec.~\ref{sec:Analysis}, the differences are only at low-energy and are already accounted for with a systematic uncertainty.

\subsection{Point-source versus finite-disk approximation}
In calculating the ICS photon flux from Eq.~\eqref{eq:emissivity-simplified}, we integrate over the angle of the incoming solar photon. In our main analysis we account for the full solid-angle integration over the Sun for the photon number density, but treat the incoming photon flux as unidirectional from the Sun's center when computing the scattering angle, rather than as a uniform surface brightness distribution across the solar disk, as in the hybrid treatment of e.g.\ Refs.~\cite{Linden:2025xom,Yang:2023res,Orlando:2008uk}. This approximation becomes exact for interaction points far from the Sun, and in particular for the outer annuli. For interaction points close to the Sun it leads to a small overestimation of the IC photon flux, which we have verified numerically to be at most $+2.2\%$, and which is negligible for this study.

\subsection{Variance inflation}
Despite profiling over the solar modulation potentials $\Phi_{e^\pm}$, the modulated background flux provides only an imperfect description of the data at lower energies, with systematic mismodeling already pointed out in Ref.~\cite{Linden:2025xom}. Given that the LIS $e^\pm$ fluxes are precisely measured and well described by the fits of Ref.~\cite{Bisschoff:2019lne} employed in this work, we attribute these residuals to limitations in the solar modulation model. Since more accurate, physically motivated models for solar modulation are not yet available, we treat this as a variance bias rather than a model bias, that is, rather than correcting the model itself, we inflate the assumed uncertainties to absorb the residual mismodeling. As already discussed in Sec.~\ref{sec:Analysis}, we introduce a relative systematic error $\epsilon_{\rm sys}$ at low energies, where $\chi^2_\nu \approx 4.5$ in the background-only fit; a constant additive systematic error would fail to capture the systematics across the wide dynamic range of the flux.

We calibrate $\epsilon_{\rm sys}$ by requiring $\chi^2_\nu \simeq 1$ in the low-energy bins under the background-only hypothesis. We calibrate on the null hypothesis rather than including a signal component, since the test statistic used to set upper limits is built to reject the null, and a signal-dependent calibration would introduce a circular dependence. This yields $\epsilon_{\rm sys} \approx 0.27$, which generically exceeds the statistical uncertainty in the low-energy bins.

To test the sensitivity of our results to this choice, we vary the systematic error over the range $0.15 < \epsilon_{\rm sys} < 0.50$, nearly halving and doubling the baseline value. The power-constrained limits shift by less than a factor of 2 across virtually the entire mass range, confirming that the constraining power is dominated by the high-energy bins where no variance inflation is applied.

\subsection{Local interstellar spectra extrapolation above 100~GeV}
The LIS electron and positron background fluxes from Ref.~\cite{Bisschoff:2019lne} are valid below a cosmic-ray $e^\pm$ energy of $\SI{100}{GeV}$. We extrapolate the fit above $\SI{100}{GeV}$ as we consider the extrapolation of the hard electron (positron) spectrum $\propto E^{-3.4}$ ($\propto E^{-3.5}$) to be more physical than a hard cutoff. This choice affects the flux in the highest energy bins within the observation window. Specifically, the computed background flux is unaffected for $\gamma$-ray energies below $\SI{10}{GeV}$, and higher energy bins are increasingly affected by additional flux of 1--100\% compared to imposing a hard unphysical cutoff.

\subsection{Energy bin and annulus subset tests}
To check internal consistency of the dataset and rule out the possibility that our limits are driven by a small number of bins with anomalous residuals, we derive constraints using complementary subsets of the data.

\paragraph{High-energy--only analysis.}
Restricting to $E_\gamma > \SI{3}{GeV}$ removes all bins affected by large solar modulation uncertainties. The resulting limits agree with the baseline to within less than a factor of 2 except for the lowest masses, confirming that the low-energy bins do not dominate the constraining power after variance inflation.

\paragraph{Inner versus outer annuli.}
We split the dataset into inner ($\theta < 15^\circ$) and outer ($\theta > 15^\circ$) annulus subsets. Both subsets yield comparable PCLs with no evidence of systematic tension between the two. The 95~\% CL\ limits at larger masses are stronger than the PCL by a factor of a few for the outer annuli while the PCL and 95~\% CL\ limits for the inner annuli are similar. The larger 95~\% CL limits for the outer annuli are likely due to a downward fluctuation in the data for the outermost annuli at high energies. Any such effects or systematics are absent in our conservative PCL analysis.

\section{All channels}\label{sec:channels}
\begin{figure*}
    \centering
    \includegraphics[scale=1.0]{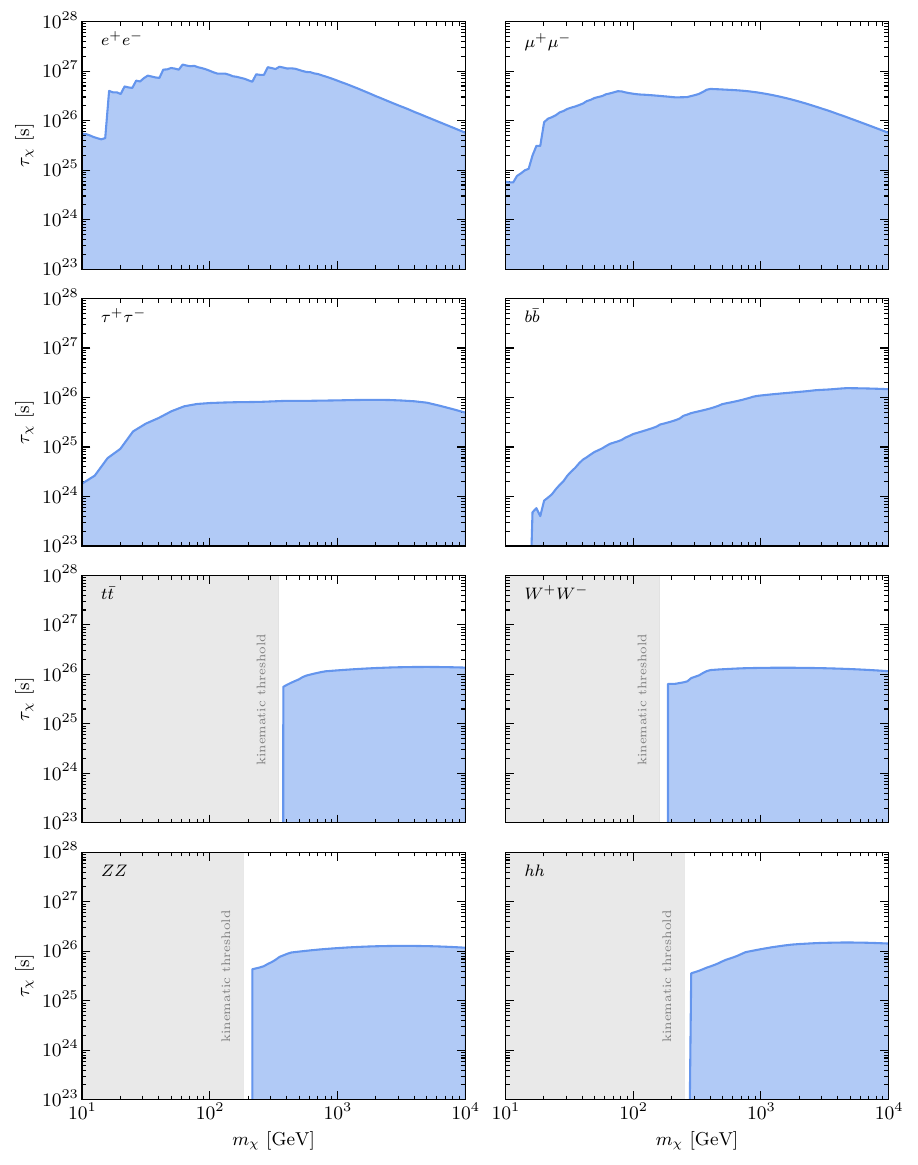}
    \caption{Power-constrained 95\% CL\ upper limits on the DM lifetime $\tau_\chi$ for all decay channels considered in this work, derived from the solar ICS $\gamma$-ray halo. Masked regions correspond to kinematic thresholds.}
    \label{fig:all-channels}
\end{figure*}

Fig.~\ref{fig:all-channels} shows the constraints on the DM lifetime for all decay channels considered in this work. The $e^+e^-$ and $\mu^+\mu^-$ results are discussed in detail in Sec.~\ref{sec:Results}. For all channels, we mask DM masses below the threshold where the PPPC4DMID spectral tables are available, which occurs at $\sim 2.3\, m_\chi$ rather than exactly at the kinematic threshold~$2 \, m_\chi$; for the $b\bar{b}$ channel this cut is more conservative, masking masses below $\SI{15}{GeV}$ due to the vanishing spectral entries near the $2m_b$ threshold in the PPPC4DMID interpolation tables.

The $\tau^+\tau^-$ channel yields constraints broadly similar to $\mu^+\mu^-$, as the secondary $e^\pm$ spectrum from the three-body $\tau$ decay is comparably soft. Quantitatively, the $\tau^+\tau^-$ limits are weaker than $\mu^+\mu^-$ by a factor of $\sim 5$ across the $\SIrange{10}{1000}{GeV}$ mass range, reflecting the larger fraction of decay energy carried away by neutrinos rather than $e^\pm$. Notably, the $\tau^+\tau^-$ constraints remain remarkably stable from $\sim\SI{100}{GeV}$ to $\sim\SI{5}{TeV}$, in contrast to the steeper high-mass falloff seen in the $e^+e^-$ and $\mu^+\mu^-$ channels. This is a consequence of the softer secondary $e^\pm$ spectrum, which reduces the impact of Klein--Nishina suppression at high energies and keeps the bulk of the upscattered signal within the \textit{Fermi}-LAT energy window to higher DM masses. The hadronic channels $b\bar{b}$, $t\bar{t}$, $W^+W^-$, $ZZ$, and $hh$ produce significantly softer secondary $e^\pm$ spectra at low masses, and consequently yield weak constraints near threshold. The limits strengthen considerably toward higher masses as the secondary $e^\pm$ spectrum hardens and shifts into the \textit{Fermi}-LAT energy range, with the $b\bar{b}$ channel showing the most pronounced mass dependence given its low threshold. Above a few hundred GeV, the hadronic and leptonic channels yield broadly comparable constraints.
We also provide heuristic fits for the DM-induced solar $\gamma$-ray spectrum (without solar modulation) in all decay channels, and model the spectrum as a centered log-parabola with a cutoff,
\begin{equation}
    E^2 \frac{\dd\Phi}{\dd E}(E,\tau_\chi,m_\chi,\theta)=A(\tau_\chi, m_\chi,\theta)\,\exp\!\left[-\beta(m_\chi)\,\ln^2\!\left(\frac{E}{E_{\mathrm{peak}}(m_\chi)}\right)\right]\,\Theta_{\mathrm{cut}}\!\left(E;m_\chi\right) \; ,
    \label{eq:fit}
\end{equation}
where the super-exponential cutoff function is
\begin{equation}
    \Theta_{\mathrm{cut}}(E;m_\chi)=\exp\!\left[-c \left(\frac{E}{E_{\mathrm{cut}}(m_\chi)}\right)^{n}\right]
\end{equation}
with typically $n \approx 2 \text{--} 3$ and $c \sim 0.1$. The cutoff energy can be estimated from Klein--Nishina suppression and the endpoint of the $e^+e^-$ spectrum at
\begin{equation}
    E_{\mathrm{cut}}(m_\chi)=\frac{\frac{4}{3}\gamma_e^2\,\epsilon_{\mathrm{ph}}}{1+4\gamma_e\epsilon_{\mathrm{ph}}/m_e},\qquad \gamma_e=\frac{m_\chi}{2m_e}.
    \label{eq:fit-peak-energy}
\end{equation}
We adopt motivated functional forms for the amplitude $A$, the fractional peak energy $a$ and the spectral curvature around the peak $\beta$.
The amplitude $A(\tau_\chi, m_\chi,\theta)$ of the $\gamma$-ray signal is a simple inverse power-law in the elongation angle $\theta$ toward the Sun as already mentioned in Sec.~\ref{sec:Analysis}.
In the Thomson regime the peak flux is constant in $m_\chi$ because the rising ICS emissivity $\propto m_\chi^2$ is exactly canceled by the $m_\chi^{-2}$ decline of the peak spectral density $\mathrm{d}\Phi/\mathrm{d}E|_{E_\mathrm{peak}} \propto 1/E_\mathrm{peak} \propto m_\chi^{-2}$, leaving $A(m_\chi) \approx \mathrm{const}$. Above the Klein--Nishina transition the cross section is suppressed, breaking this cancellation and causing the amplitude $A$ to fall. A natural form is therefore
\begin{equation}
    A(\tau_\chi, m_\chi, \theta) = A_0 \times \frac{(m_\chi/m_{\rm ref})^{2-p_A}}{1 + (m_\chi/m_*)^{\gamma_{A}}} \times \left( \frac{\SI{e25}{s}}{\tau_\chi} \right) \times \left(\frac{\theta}{10^\circ}\right)^{-q_{\theta}} , \quad m_* \sim m_e^2/(2\varepsilon_\mathrm{ph}) ,
    \label{eq:fit-amplitude} 
\end{equation}
where $q_\theta \approx \num{0.9}$ and $m_{\rm ref} = \SI{1}{GeV}$. $m_{*} \approx \SI{330}{GeV}$ is the Klein--Nishina transition mass set by the solar photon energy $\varepsilon_\mathrm{ph}$. In Eqs.~\eqref{eq:fit-peak-energy} and \eqref{eq:fit-amplitude} we use $\varepsilon_{\rm ph} \approx \SI{0.4}{eV}$.
We model the fractional peak energy $a = E_{\rm peak} / E_{\rm cut}$ by a broken power law
\begin{equation}
    a(m_\chi) = a_0 \left[ \left(\frac{m_\chi}{m_0}\right)^{-\alpha_1 \delta} + \left(\frac{m_\chi}{m_0}\right)^{-\alpha_2 \delta} \right]^{1/\delta}\; .
    \label{eq:fit-fractional-peak-energy}
\end{equation}
For the coefficient of the quadratic log-term, we fit a simple power law for the dependence on the DM mass,
\begin{equation}
    \beta(m_\chi)=b_\infty+b_0\left(\frac{m_\chi}{100\,\mathrm{GeV}}\right)^{-b_1}\; .
    \label{eq:fit-curvature}
\end{equation}
We perform fits independently for each decay channel and list the fitted parameters in Tab.~\ref{tab:fits}. Note that the spectral curvature at the peak is a non-trivial function for channels with typically longer decay chains, namely $b\bar{b}$, $t\bar{t}$, $W^+W^-$, $ZZ$ and $hh$ due to the injected $e^+ e^-$ spectrum being itself a broad power-law. We therefore refrain from fitting $\beta(m)$ for these decay channels.
\begin{table*}[t]
    \centering
    \renewcommand{\arraystretch}{1.2}
    \begin{tabular}{l @{\hspace{2em}} S[table-format=1.2e-1] S[table-format=1.2] S[table-format=1.2] @{\hspace{2em}} S[table-format=1.3] S[table-format=2.2] S[table-format=1.2] S[table-format=1.1] l @{\hspace{2em}} S[table-format=1.3] S[table-format=1.2] S[table-format=1.2]}
        \toprule
        Channel & {$A_{0}$} & {$p_A$} & {$\gamma_{A}$} & {$a_0$} & {$\alpha_1$} & {$\alpha_2$} & {$\delta$} & {$m_a$} & {$b_{\infty}$} & {$b_0$} & {$b_{1}$} \\
        & {[\si{GeV\,cm^{-2}\,s^{-1}\,sr^{-1}}]} & & & & & & & {[GeV]} & & & \\
\midrule
        $e^+ e^-$       & 3.06e-07 & 1.72 & 1.50 & 0.5   & 0    & 0    & 1   & 300 & 0.034 & 0.10 & 0.58 \\
        $\mu^+ \mu^-$   & 2.35e-07 & 1.81 & 0.96 & 0.5   & 0.6  & 0.0  & 5   & 30  & 0.003 & 0.08 & 0.50 \\
        $\tau^+ \tau^-$ & 4.52e-08 & 1.70 & 0.90 & 0.105 & 0.75 & 0.0  & 4   & 95  & 0     & 0.12 & 0.22 \\[3pt]
        $b \bar{b}$     & 8.47e-09 & 1.44 & 0.81 & 0.009 & 2.0  & 0.34 & 5   & 90  & \multicolumn{3}{c}{---} \\
        $t \bar{t}$     & 3.28e-09 & 1.31 & 0.90 & 0.013 & 0.42 & 0.42 & 1   & 10  & \multicolumn{3}{c}{---} \\
        $W^+ W^-$       & 1.25e-08 & 1.54 & 0.73 & 0.022 & 10.0 & 0.7  & 0.6 & 200 & \multicolumn{3}{c}{---} \\
        $ZZ$            & 2.40e-09 & 1.28 & 0.95 & 0.007 & 0.45 & 0.45 & 1   & 100 & \multicolumn{3}{c}{---} \\
        $hh$            & 8.70e-10 & 1.09 & 1.11 & 0.003 & 0.25 & 0.25 & 1   & 100 & \multicolumn{3}{c}{---} \\
        \bottomrule
    \end{tabular}
    \caption{Best-fit values for the parameter functions for the amplitude $A$ (namely $A_0$, $p_A$, $\gamma_A$ in Eq.~\eqref{eq:fit-amplitude}), the ratio of the peak energy versus cutoff energy $a$ (namely $a_0, \alpha_1, \alpha_2, \delta, m_a$  in Eq.~\eqref{eq:fit-fractional-peak-energy}), and the spectral curvature at the peak $\beta$ (namely $b_\infty, b_0, b_1$ in Eq.~\eqref{eq:fit-curvature}) for all decay channels. The hadronic and bosonic channels ($b\bar{b}$, $t\bar{t}$, $W^+ W^-$, $ZZ$, $hh$) involve longer decay chains into $e^\pm$ with non-trivial spectral shapes; the curvature parameters $b_\infty$, $b_0$, $b_1$ are therefore not fitted for these channels.}
    \label{tab:fits}
\end{table*}
\end{document}